\documentclass[preprint]{elsarticle}
\biboptions{sort&compress}
\RequirePackage{amsmath}
\newcommand{\dfr}[2]{\frac {\displaystyle #1}{\displaystyle #2}}
\begin{document}
\begin{frontmatter}
\title{About causes of slow relaxation of melted intermetallic alloys}

\author[mymainaddress1,mymainaddress2]{V.G. Lebedev}
\author[mymainaddress1,mymainaddress2]{K.Y. Shklyaev}
\author[mymainaddress1,mymainaddress2]{S.G. Menshikova}
\author[mymainaddress3,mymainaddress4]{M.G. Vasin}

\corref{mycorrespondingauthor}
\ead{dr\_vasin@mail.ru}
\address[mymainaddress1]{Udmurt Research Center of Ural Branch of Russian Academy of Sciences, Izhevsk 426000, Russia}
\address[mymainaddress2]{Udmurt State University, Izhevsk 426000, Russia}
\address[mymainaddress3]{Vereshchagin Institute of High Pressure Physics, Russian Academy of Sciences,  Moscow 108840, Russia}
\address[mymainaddress4]{Institute of Metallurgy of Ural Branch of Russian Academy of Sciences, Ekaterinburg, Russia}

\begin{abstract}
Ascertainment of the nature of the slow relaxation processes observed after melting in glass-forming eutectic melts is the subject of this work. We claim that the diffusion processes nonlinearity in heterogeneous melt with inclusions of refractory stoichiometry is the origin of this phenomenon. The cause for this  nonlinearity is the thermodynamic instability similar to one taking place at spinodal decomposition, and indispensable condition is the initially non-homogenous. For confirmation of our devotes, we consider the model of liquid solution of a binary system, which evolution described by the Cahn--Hilliard equation with combined Gibbs potential assuming the presence of remains after melting stoichiometric phase. Exemplified by the Al--Y and Al--Yb alloys, using Gibbs potentials from standard database we show that subject to initial heterogeneity in these systems the instability can develop  leading to the slow relaxation  processes, and determine the regions of this instability in the phase diagrams.
\end{abstract}

\begin{keyword}
intermetallic alloys, relaxation, melts, viscosity
\end{keyword}
\end{frontmatter}

\section{Introduction}

\begin{figure}
\centering
\includegraphics[scale=0.8]{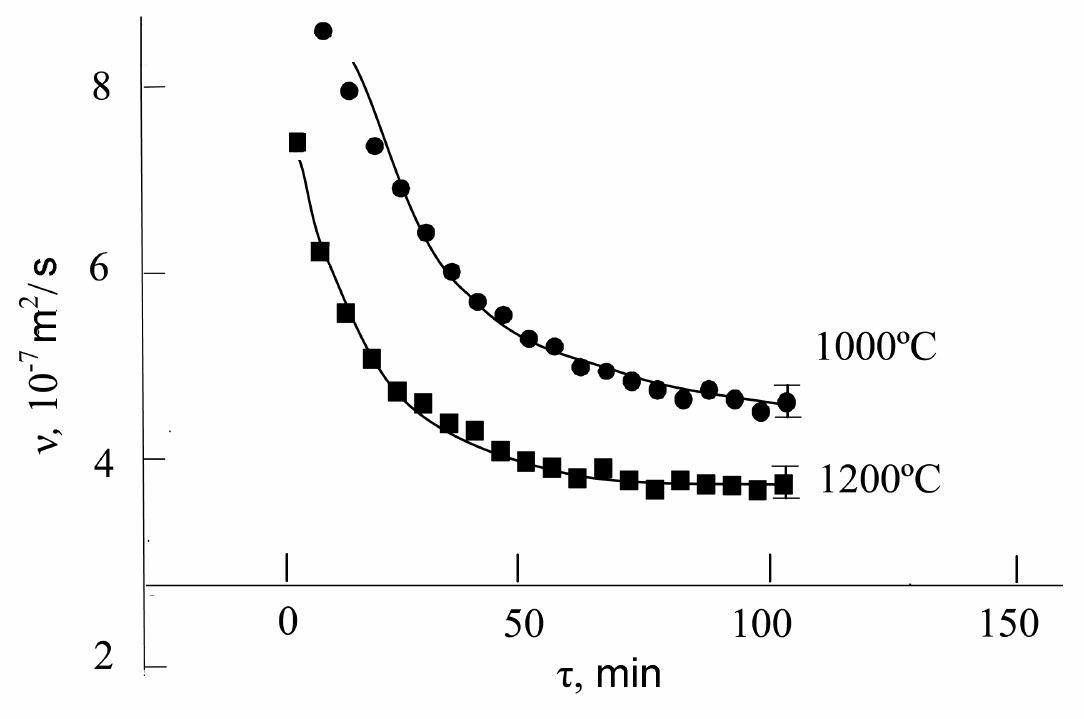}
\caption{The time dependencies of the Al$_{90}$Y$_{10}$ melt viscosity at 1000$^{\circ}$C and 1200$^{\circ}$C \cite{4}.}
\label{fig1}
\end{figure}

This work is devoted to the study of physical processes observed in some intermetallic melts, which nature remains completely unexplained. Namely, we consider the phenomenon of slow relaxation of some intermetallic melts after melting, when the relaxation time reaches few hours~\cite{Son,1,2}. As an example of the slow relaxation phenomenon, on the Fig.\,\ref{fig1} shows the characteristic experimentally observed dependence of the melt viscosity of Al--Y on time after melting.
This relaxation is sometimes non-monotonic in time \cite{2, 3, 4, 5}. In this case, for a certain period of time after melting, the melt viscosity decreases exponentially, but at some point it suddenly begins to grow, reaching a local maximum, and then returns to the normal exponentially decreasing mode. Usually, these effects are explained as remelting processes, in which slow melting of refractory solid phase fragments takes place. However, the linear diffusion theory can not to explain these slow relaxation processes.  This theory gives the estimation of order of some seconds for the relaxation time of melt. It is very far from the relaxation time values observed in the experiments, which are of the order of some hours.

The slow relaxation is observed at investigations of structure-sensitive melts properties: viscosity, density or resistivity. Significant progress in understanding of this phenomenon was achieved at the end of the 20-th century. It was found that the slowdown is accompanied by a long period of inhomogeneity retention in the form of metastable microemulsion, in which droplets of 10--100 angstroms in size exist for a long time. This conclusion was also supported by the direct structure investigations \cite{S1,S2}.
As we noted above, in the Al--Y, Al--La and Al--Ce melts not only slow relaxation but also non-monotonic in time processes are observed \cite{4,VMI}. Similar slow relaxation is observed in other aluminium-based intermetallic melts (Al--Ce, Al--Sm) \cite{BLS}.
In \cite{Gasser} in the stable resistivity investigation it was founded, that the melt of Bi-In remains inhomogenius a long time.
To obtain a macroscopic homogeneous liquid alloy, it was necessary to wait several days and to heat to 700 C$^{\circ}$ above the melting point  \cite{Gasser}.

Recently, we suggested the theoretical description of this phenomenon, in terms of well known Cahn--Hilliard theory~\cite{VMI,LV}, using the combined Gibbs potential assuming the presence of remains after melting stoichiometric phase. In this paper, we develop this approach, considering initially the system as a system with impurity concentration fluctuation. Stating briefly, we suppose all these kinetic phenomena are the result of the physical system non-linearity appearing because of its initial heterogeneity. The non-linearity is the result of the atoms of dissolved component are arrested in the regions of its high concentration long time, since in these regions the stoichiometric structure remains to be energetically more favoured.

\begin{figure}
\centering
\includegraphics[width=0.8\textwidth]{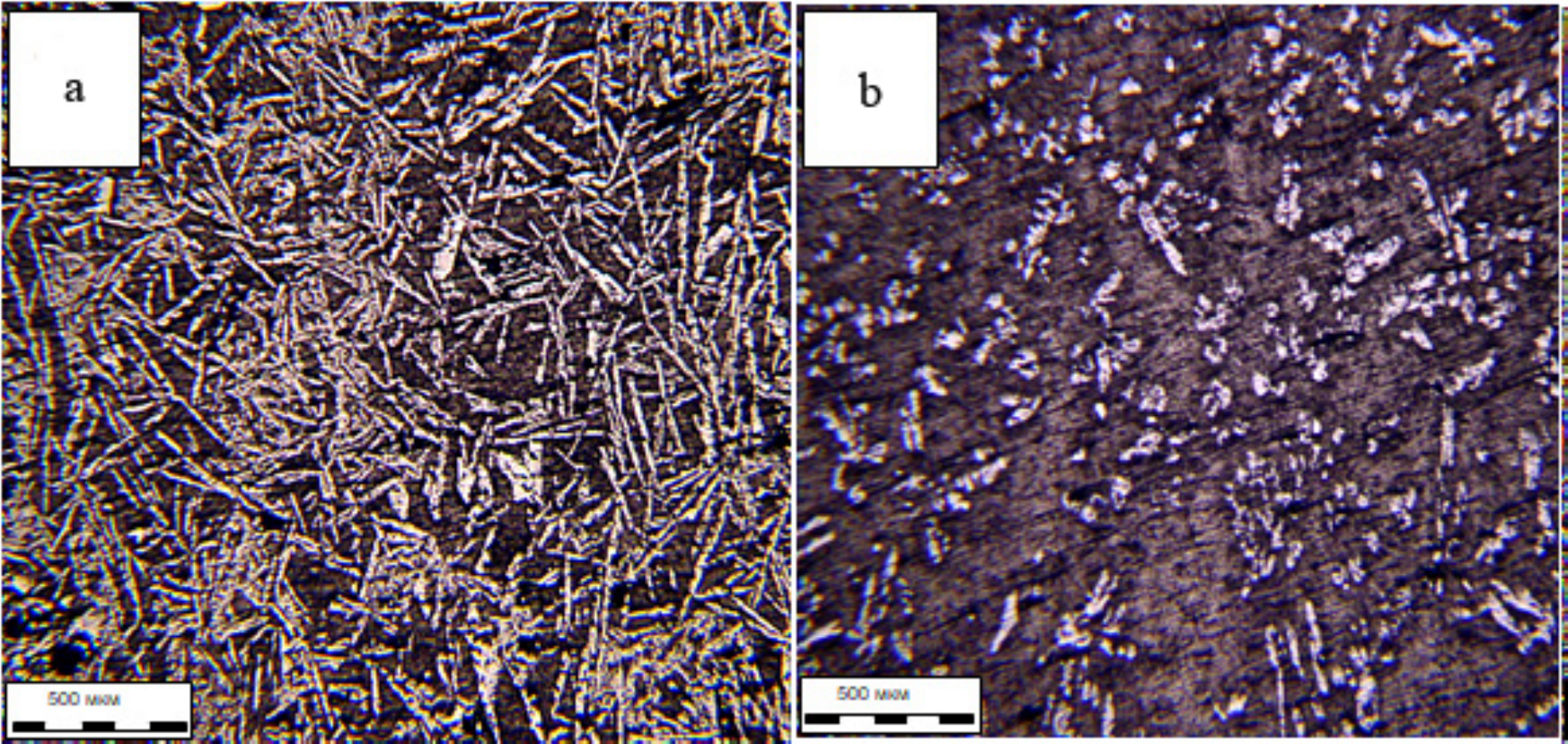}
\caption{Ligatures of initial samples a) Al--Y10\%, b) Al--Y5\% (from \cite{RMET}). In the structure one can see the inclusions of stoichiometric Al$_3$Y compounds which characteristic size $\sim 10^{-5}$ m.}
\label{LIG}
\end{figure}

Below, we consider the relaxation of a binary intermetallic alloy melt obtained by melting an initial solid sample whose structure is a solid solution containing inhomogeneities in the form of inclusions of a stoichiometric phase. In particular, for the considered Al--Y this initial inhomogeneities is presented in Fig.\,\ref{LIG}.

\section{Description of the kinetics of relaxation process}

To describe the kinetics of relaxation process of the continuous medium with non-uniform impurity distribution, let use the Cahn--Hilliard equation \cite{CH} for solute concentration $c({\bf r},t)$, which in general case has is written as follows:
\begin{align}
\label{CHL1}
\frac{\partial c}{\partial t}=M_D{\nabla}^2\Big(\frac{\partial{\cal F}}{\partial c}\Big).
\end{align}
Here $M_D$ ($M_D>0$) is the transport coefficient, and ${\cal F}$ is the density of free energy, which has the following form:
\begin{align}
\label{CHF}
{\cal F}=f_0+f(c)+\frac12\varepsilon^2(\nabla c)^2.
\end{align}
The two first terms in this expression are the free energy of the solution, and the third one $\varepsilon^2(\nabla c)^2$ ($\varepsilon^2>0$) describes the non-homogeneity contribution of the solution to the free energy. In addition, we assume that the molar volume is independent of composition.

Note,  if $f(c)$ potential has only one minimum, the equation (\ref{CHL1}) describes the linear diffusion, and the characteristic dispersion relation is written as follows: $\omega=-k^2(D+M_D\varepsilon^2k^2)$ \cite{SS}, where $D=M_D{\partial^2f(c)}/{\partial c^2}$ is the diffusion coefficient. If the $f(c)$ is non-linear and contains two or more minima, then the system kinetics becomes not so simple. The best example of like these processes is the spinodal decomposition in liquids. It takes place when $f(c)$ has two minima. In this case, at some concentration area, the $f(c)$ function has a negative curvature. As a result, in this area the diffusion coefficient $D<0$, that leads to uphill diffusion, when the fluctuation induced heterogeneities do not decrease, but growth. This instability in combination with the solute mass preserve condition leads to ``worm-like'' structure in the sample.  The processes of relaxation at the spinodal decomposition are very slow, since in this case, the effective diffusion coefficient becomes very small.

We suppose that slow relaxation processes in the considered by us melts have the physical cause similar to one for the relaxation slowing down in the spinodal decomposition. The cause of the non-linearity in this case is the existence of stoichiometric compounds in the alloys.
From numerous of experimental observations, it is known that in solid state the intermetallic alloys are non-homogenous in structure. They are often a solid solution structure with inclusions of stoichiometric compounds. After melting, it leads to presence of the initial strong heterogeneity of the melt. These non-homogenous constitute the local areas in which the solute concentration many differences to the sample-average.
It results, the Gibbs potential of a heterogeneous melt has more complex form than the homogenous solution Gibbs potential. This potential is the sum of the liquid solution Gibbs potential and the Gibbs energy of the stoichiometric compounds, leftover after melting of the initial solid sample. We suppose just this is the origin of the non-linearity, and, as a result, the slow kinetics of relaxation of the considered systems.

\section{Relaxation in the melt of binary alloy with stoichiometry}

We consider the binary melt model, in which $c$ is the average solute concentration. We suggest right after melting, this melt non-homogeneous because non-homogeneity of initial solid sample. The non-homogeneity in the melt makes oneself evident in the existence in it structure of the fluctuating clusters of the stoichiometric phase. As a result, the concentration of impurity in some local area can be not equal to average one, $c({\bf r},\,t)\neq c$, and depends on the phase state of both this area and around areas.

Let us introduce the scalar field $\varphi $ for description of the phase state of the non-homogeneous melt. If in some volume unit, the stoichiometry arises, then $\varphi $ ($0\le \varphi\le 1$)  is the stoichiometry share, and the share of the liquid solution in this volume unit is $1-\varphi $.  Thus, if we assume that $\varphi=1$ in the solid phase, then this field is zero in the liquid phase. The amount of the solute in the volume unit is the sum of the impure atoms in the solid, $c_c$, and in the liquid, $c_l({\bf r})$, phases:
\begin{align}\label{CL}
c({\bf r},\,t)=(1-\varphi({\bf r},\,t))c_l({\bf r},\,t))+\varphi({\bf r},\,t) c_c  \qquad \left( \varphi=\dfr{c-c_l}{c_c-c_l}\right),
\end{align}
The Gibbs energy of this system can be presented as the sum of the Gibbs potentials of the solid and liquid parts:
\begin{align}
\label{CL}
f(c_l,\,\varphi)=G_c\varphi+G_l(c_l)(1-\varphi),
\end{align}
where $G_l(c_l)$ is the Gibbs potential of the liquid Al--Y solution, and $G_c$ is the Gibbs potential of the stoichiometric phase.

For simplicity, we will neglect a volume change during phase transformation. In the spirit of the quasi-equilibrium theory \cite{Flem}, the $\varphi $-field corresponds only to the volume share stoichiometric phase, and does not describe the  interface.
Therefore, we suppose the contribution of only the gradient of concentration $c({\bf r})$ to the melt free energy. Thus, the molar free energy function is written as follows:
\begin{align*}
  {\cal F}=f_0+f(c_l,\,\varphi)+\frac12\varepsilon^2(\nabla c)^2.
\end{align*}

\subsection{Relaxation of homogeneous system}

The important thing is that the local molar solute concentration, $c_l({\bf r})$, can change both due to the solid phase fraction vary, and because of the atom diffusion in the liquids. While the concentration $c({\bf r})$ vary only due to diffusion fluxes ${\bf J}_D$ in liquid phase, which fraction is $(1 - \varphi)$. Therefore
\begin{align*}
\partial_t c({\bf r})=-(1-\varphi)\nabla\cdot{\bf J}_D.
\end{align*}
The diffusion fluxes are defined from the minimisation of the system Gibbs energy. One can show in our case these fluxes are proportional to the  molar chemical potential gradients, $\mu ={\partial {\cal F}}/{\partial c_l}$:
\begin{align*}
{\bf J}_D=-M_D\nabla \dfr{\partial {\cal F}}{\partial c_l}.
\end{align*}
As a result
\begin{align*}
\partial_t c({\bf r})=(1-\varphi)\nabla\left( M_D\nabla \dfr{\partial {\cal F}}{\partial c_l}\right).
\end{align*}
Taking into account that $M_D$ is constant, and $\varphi\approx 0$ we rewrite this expression as follows:
\begin{align*}
\partial_t c({\bf r})\approx M_D\nabla^2 \dfr{\partial {\cal F}}{\partial c_l}= M_D\nabla^2\left(\left.\dfr{\partial f}{\partial c_l}\right|_{c_l=c}-\nabla^2c\right).
\end{align*}

If the system is in equilibrium, then $\partial_t c=0$.
Hence, we come to well known in the general thermodynamic, condition of equilibrium:
\begin{align*}
  \left.\dfr{\partial f}{\partial c_l}\right|_{c_l=c=c^*}=0, \left.\quad \nabla c_l\right|_{c_l=c=c^*}=0,
\end{align*}
The derivation $f$ over $c_l$ can be rewritten as follows:
\begin{align*}
\dfr{\partial f}{\partial c_l}=\dfr{c_c-c}{c_c-c_l}\left[\dfr{\partial G_l}{\partial c_l}-\dfr{G_c-G_l}{c_c-c_l}\right].
\end{align*}
Taking into account that close to the liquidus
the system is liquid, i.e. $c=c_l=c^*$ and $\varphi \ll 1$, we find that the liquidus position corresponds to the point $(c^*,\,T)$ on the phase diagram for which
\begin{align}
  \left.\dfr{\partial G_l}{\partial c_l}\right|_{c_l=c^*}-\dfr{G_c-G_l}{c_c-c^*}=0.
\label{BIN}
\end{align}
One can see this is a well-known expression defining the liquidus line position in the phase diagram.

\subsection{Relaxation of initially heterogeneous system}

For simplicity, we suggest $\nabla \left({\partial f}/{\partial c_l }\right)=0$. This is a strong assumption. However, in case of strongly disordered system, on high space scales this suggestion can be applicable. Also, in this case, the series expansion of the free energy, $f$, over $\nabla c({\bf r})$ can be confined by the quadratic term. Thus, the system evolution is described by the following expression:
\begin{align}
\label{EQQ}
\partial_t c\approx M_D\left(\left.\dfr{\partial^2f}{\partial c\,\partial c_l}\right|_{c_l=c}\nabla^2c-\varepsilon^2\nabla^4c\right),
\end{align}
where
\begin{align*}
\dfr{\partial^2 f}{\partial c\,\partial c_l}=-\dfr{1}{c_c-c_l}\left[\dfr{\partial G_l}{\partial c_l}+\dfr{G_l-G_c}{c_c-c_l}\right].
\end{align*}
This naturally leads to the critical slowing of the relaxation processes near the liquidus line. The boundaries of the concentration region in which this slowing can be observed are determined both by the thermodynamic potential of the system and by the initial conditions.

Let us consider initially heterogeneous system with concentration fluctuation, $\delta c$, around the equilibrium average concentration $c_l=c$. Then the system's molar free energy function has the following form
\begin{gather*}
f(T,\,c_l+\delta c)=f(T,\,c_l)+\dfr12\dfr{\partial^2f(T,\,c_l)}{\partial c_l^2}\langle\delta c^2\rangle+\dfr1{4!}\dfr{\partial^4f(T,\,c_l)}{\partial c_l^4}\langle \delta c^4\rangle+\ldots .
\end{gather*}
In this expression only terms with even degrees of $\delta c$ are present. This is because the deviation can be both positive and negative sign, therefore, when averaging, the terms with odd degrees are equal to zero.
Thus, the evolution equation (\ref{EQQ}) is rewritten as follows:
\begin{align}\label{EQQ2}
\partial_t \delta c\approx M_D\left(K(T,\,c,\,\delta c)\nabla^2\delta c-\varepsilon^2\nabla^4\delta c\right),
\end{align}
where $K(T,\,c,\,\delta c)$ function has the following form:
\begin{align}\label{EQQ3}
\displaystyle K(T,\,c,\,\delta c)=\dfr{\partial}{\partial c}\left.\dfr{\partial f(T,\,c_l)}{\partial c_l}\right|_{c_l=c}+\dfr12\dfr{\partial}{\partial c}\left.\dfr{\partial^3f(T,\,c_l)}{\partial c_l^3}\right|_{c_l=c}\langle\delta c^2\rangle\\
\displaystyle +\dfr1{4!}\dfr{\partial}{\partial c}\left.\dfr{\partial^5f(T,\,c_l)}{\partial c_l^5}\right|_{c_l=c}\langle\delta c^4\rangle+\ldots,
\end{align}
and
\begin{align}\label{EQQ4}
\dfr{\partial}{\partial c}\dfr{\partial^{n} f}{\partial c_l^n}=
-\dfr{1}{c_c-c_l}\left(\sum\limits_{i=1}^n\dfr{n!/i!}{(c_c-c_l)^{n-i}}\dfr{\partial^i G_l}{\partial c_l^i}+\dfr{n!}{(c_c-c_l)^{n-1}}\dfr{G_l-G_c}{c_c-c_l}\right).
\end{align}

From the above one can see the system is stable when $K(T,\,c,\,\delta c)>0$.  This is if $\delta c=0$, and only the first term in (\ref{EQQ3}) is relevant. However, in the presence of concentration fluctuations the rest terms also become relevant, and can give contribution with negative sign. Then the magnitude of the fluctuation of the concentration of the second component relative to its equilibrium mean value significantly affects the $K$, and the positive sign of the first series term in (\ref{EQQ3}) is no longer the criterion of stability. If at some fluctuation amplitude, the $K$ becomes negative, then the conditions for this fluctuation growth arise. In spite of the fact that this growth is limited by the condition of concentration continuity, the system dynamics becomes nonlinear. It is the same as in the case of spinodal decay, the only difference being that the relaxation process occurs in inverse way: from initial  non-homogenous state to full homogeneity.

Using (\ref{EQQ2}), and the Gibbs potentials of liquid solution, $G_l(c_l)$, and stoichiometric phase we can take into account the effects of this non-linearity, and determine the conditions of stability/instability of the system for given initial concentration deviation from its mean value. Thus, the boundary of the region in which instability due to $\delta c$ fluctuation is possible is determined from the condition $K=0$.

\section{Determination of the non-linear relaxation area in phase diagram}

Using the above, one can determine in the phase diagrams of intermetallic melts the areas in which this non-linear relaxation is possible. For example, we consider relaxation of Al--Y-based melts, which earlier was already discussed in \cite{VMI, LV}. The initial solid samples using in the considering experiments contained stoichiometric inclusions of Al$_3$Y and Al$_2$Y with characteristic size order $10^{-5}$ m (see Fig.\,\ref{LIG}). From the point of view of the linear theory of diffusion the characteristic dissolution time of these inhomogeneities should be $10^{-2}$ s, that is many fewer than the relaxation time, which is observed in the experiment: $\tau\sim ~10^{4}$ s \cite{VMI}. As we have assumed above, the nature of this unusual phenomenon can be explained by the thermodynamic instability of heterogeneous melts.

\begin{figure}
\centering
\includegraphics[width=0.6\textwidth]{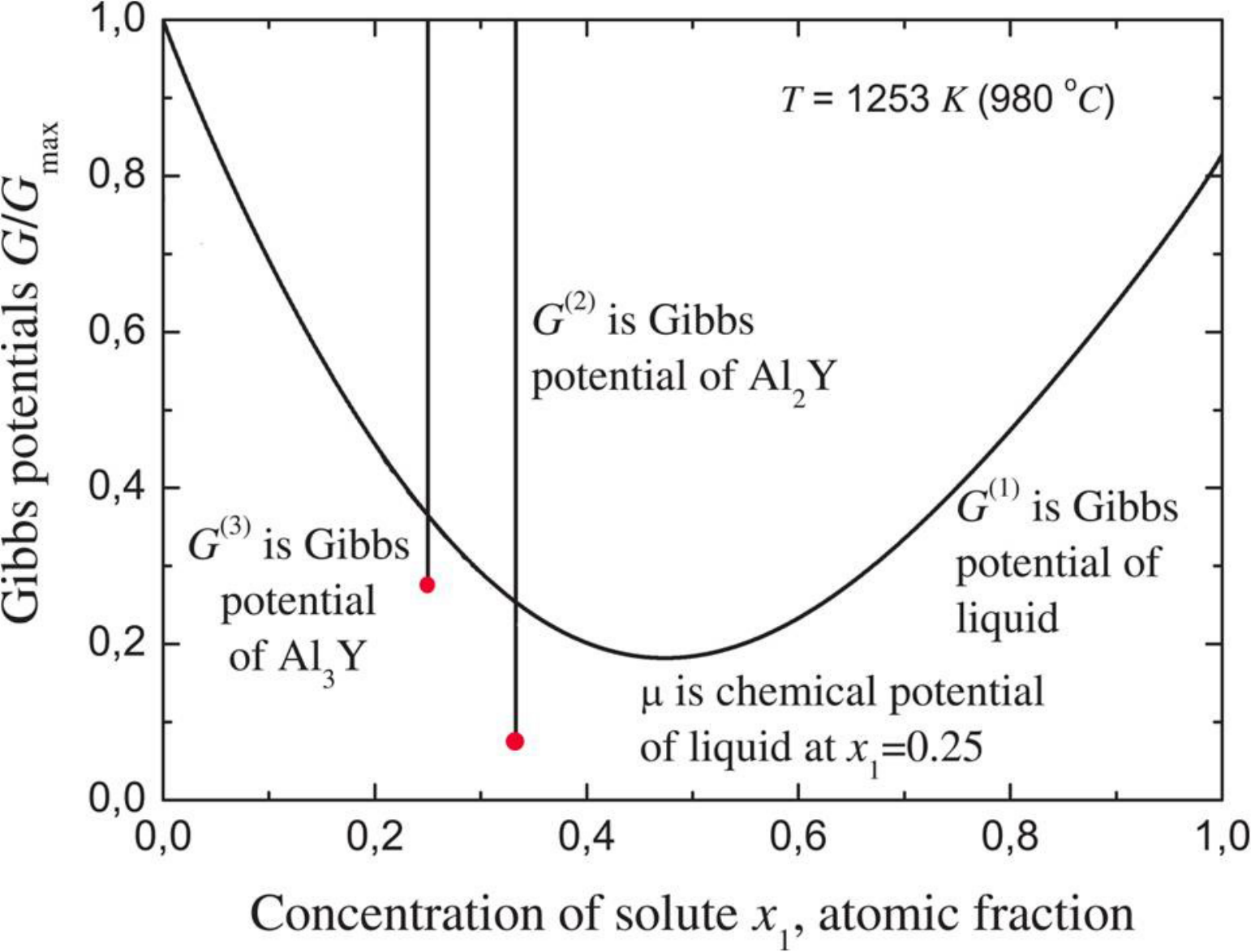}
\caption{Gibbs potentials of the liquid phase and two stoichiometric phases, Al$_3$Y and Al$_2$Y, in Al--Y alloy at the temperature $T=980^{\circ}$C from the Computational Phase Diagram Database of Japanese National Institute for Materials Science \cite{nims}.}
\label{fig3}
\end{figure}

\begin{figure}
   \centering
   \includegraphics[scale=0.8]{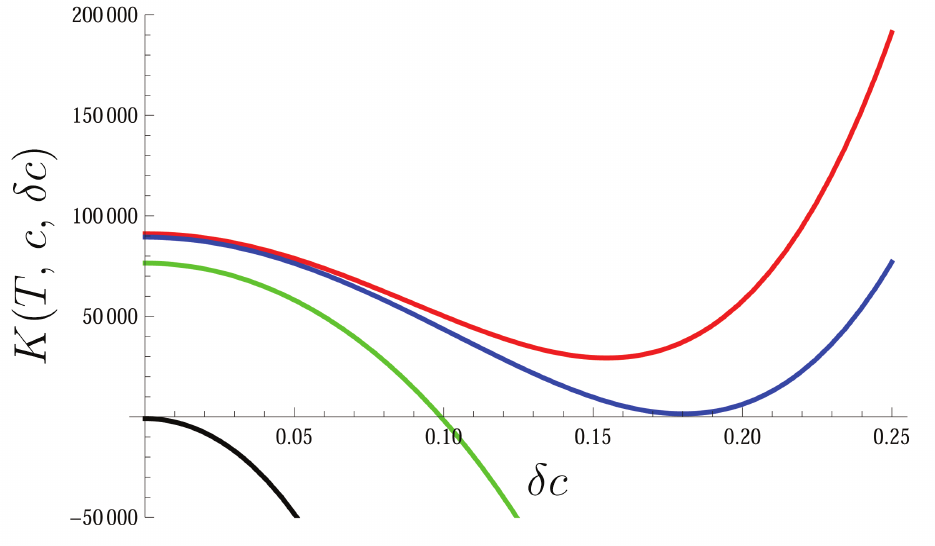}
   \caption{The calculated concentration fluctuation dependence of the $K$ function of the Al--Y melt at $T=1273$ K around $c=0.07$ (red line), $c=0.0712$ (blue line), $c=0.08$ (green line), and $c=0.12$ (black line). One can see that at the equilibrium concentration exceeding $c=0.0712$ the $K$ becomes negative in some interval of the concentration deviations from equilibrium value. At $c=0.12$ the liquid Al--Y solution becomes non-equilibrium.}
   \label{F2}
\end{figure}

Based on the above theoretical arguments, we can estimate the conditions of thermodynamic instability appearing. Using the known from~\cite{nims} (Fig.\,\ref{fig3}) Gibbs energy functions for the liquid solution of yttrium in aluminum, and Al--Y stoichiometric compounds. In principle, using (\ref{EQQ4}), one can calculate any term of the (\ref{EQQ3}) series. However, we will limit ourselves further to only the first three ones, since this is quite enough for the necessary accuracy. The calculated $K$ function for the considered binary melt is shown in Fig.\,\ref{F2}, in which one can see that there is the critical deviation of concentration from equilibrium value, at which the system relaxation dynamics becomes instable ($K<0$).
Note, again, that the origin of this instability needs the structural heterogeneity of the initial melt. In this case, one should hardly expect a significant increase in the initial inhomogeneities, or an infinitely long relaxation. However, the diffusion in this case becomes essentially nonlinear, and the relaxation time is no longer determined by the diffusion coefficient of a single atom in a given liquid. For this reason, the relaxation processes of homogenization become much slower.

For comparison with the available experimental data, below we will perform a quantitative estimation of the initial non-homogeneities sizes, that lead to the relaxation slowing, using the expressions above obtained.
For description of nucleation kinetics, one can use the dynamical structure factor $S({\bf k},t)$, which, as it is known, is proportional to the time-depended of the pair correlation function of concentration \cite{SS},
\begin{align*}
S({\bf k},\,t)=C({\bf k},0)e^{-A({\bf k})t/2}=\int\left[c_l({\bf r},\,t)-c\right]e^{-i{\bf kr}}\mathrm{d}{\bf r},
\end{align*}
and can be expressed as $S({\bf k},\,t)=\langle C^2({\bf k},\,t)\rangle\propto\exp(-A({\bf k})t)$, where the $A({\bf k})$ is the amplification rate. The amplification rate can be estimated from the dispersion relation, which ensues from the Cahn--Hilliard equation (\ref{EQQ2}):
\begin{align*}
A({\bf k}) = M_D{\bf k}^2\left(K(T,\,c,\,\delta c)+\varepsilon^2{\bf k}^2\right).
\end{align*}
One can see that the relaxation slows when the amplification rate approach to zero. It is possible when $K$ becomes negative and the characteristic space scale of initial inhomogeneities is
$\xi\sim\sqrt{-\varepsilon^2/K}$. Thus, we have got the important conclusion that the initial heterogeneity scale leading to slow relaxation phenomenon is inversely proportional to the square root of the thermodynamic force of crystal formation in the area with high concentration of the second component.

From Fig.\,\ref{F2} one can see, that in the Al--Y melt with mean yttrium concentration $c\approx 8$ \% at the temperature $T=1273$ K the heterogeneities with yttrium concentration $c+\delta c\sim 19 - 21$ \%  can be stable since $K$ is negative  and $|-K|\sim 10^4 - 10^6$ J/mol.  Usually, the magnitude order of $\varepsilon^2$ is $10^{-7}$ J$\cdot$m$^2$/mol. Thus, we can estimate the minimal size of the relatively stable heterogeneity, which is $\xi\sim 10^{-5} - 10^{-6}$ m.
The larger the size of the heterogeneity with a higher content of the second component, the slower its relaxation.
This allows to explain the observed slow relaxation, since the examinations of initial solid samples give the scale of the stoichiometric inclusions, which in several orders of magnitude exceeds the estimation of the minimal heterogeneity size which is necessary for the slow relaxation manifestation (see Fig.\,\ref{LIG}).

Thus, the long-lived concentration inhomogeneities can exist in the melt under the condition $K<0$. And the condition $K=0$ determines the boundary of the temperature-concentration region of the existence of such non-linear (slow) relaxation in the phase diagram.
For example, in the Al--Y melt at temperature $T=1273$ K, this boundary corresponds to the concentration $c=0.0712$ at which the $K(T,\,c,\,\delta c)$ function touches the abscissa axis on the $K$ vs $\delta c$ graph (blue line in Fig.\,\ref{F2}).

Just as in the case of the description of spinodal decay \cite{SS}, the above analysis is applicable only to the initial stages of relaxation. In the considered inhomogeneous system, a negative value of $K$ in some local region does not indicate an infinite growth of this region, since it can be part of a larger region for which $K>0$. Eventually, the system comes to an equilibrium state with a uniform distribution of the second component. If a more precise analysis is needed, it is necessary to analyze the resulting nonlinear differential equation with all the appropriate initial and boundary conditions.

\subsection{Estimation of non-linear (slow) relaxation area in phase diagrams}

\begin{figure}
   \centering
   \includegraphics[scale=1.2]{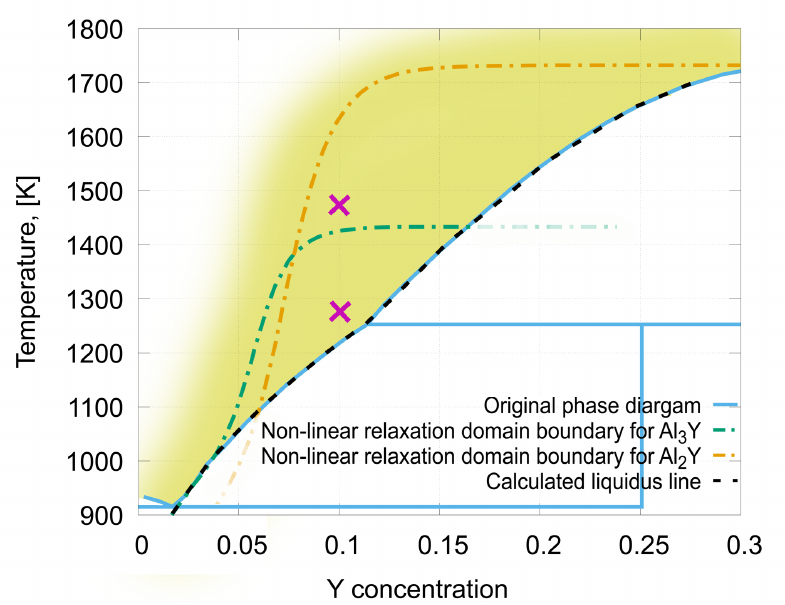}
   \caption{Al--Y phase diagram with highlighted (non-linear) slow relaxation area in the liquid phase.}
   \label{F4}
\end{figure}

The Computational Phase Diagram Database of Japanese National Institute for Materials Science databases for binary systems were used for the calculation \cite{nims}. From these bases the standard Gibbs potentials reconstructing the phase diagrams of the described systems were taken: Foremost the expression for the concentration dependence of the total molar Gibbs energy $G_l(c)$ at a given temperature $T$, as well as the Gibbs energy for the stoichiometric phases $G_c$ and the molar fraction of impurity atoms $c_c$ in the stoichiometric phase were found;
for verification, using the data obtained and the condition (\ref{BIN}), we found the liquidus line, which coincided with the experimental one;
substituting the parameters in (\ref{EQQ3}) and (\ref{EQQ4}) allows  the boundary of the nonlinear relaxation region in the phase diagram.

\begin{figure}
   \centering
   \includegraphics[scale=1.2]{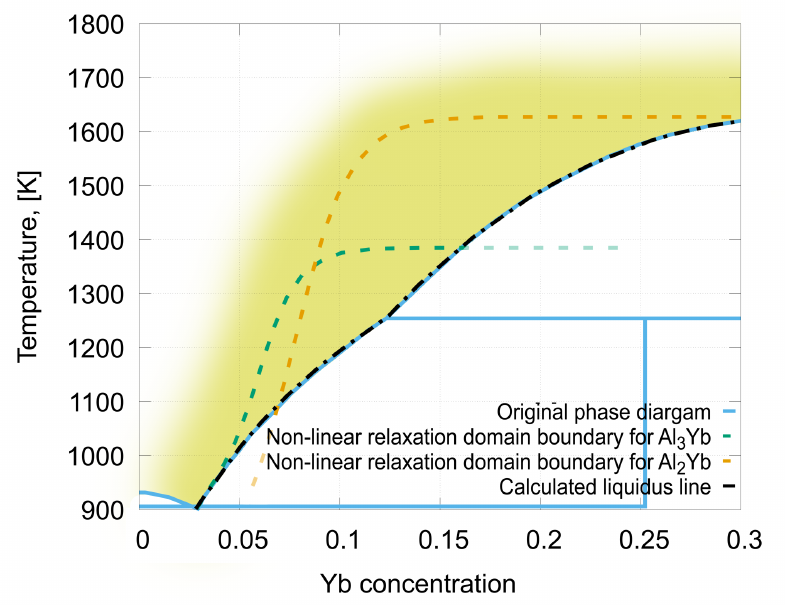}
   \caption{Al--Yb phase diagram with highlighted (non-linear) slow relaxation area in the liquid phase.}
   \label{F5}
\end{figure}

We have done this procedure for the Al--Y and Al--Yb melts. The results are shown in Fig.\,\ref{F4} and  Fig.\,\ref{F5}. From the figures, one can see that at a given temperature, the area of non-linear system behavior can spread out in a wide concentration interval on the left of liquidus line.  However, it should be understood that the lines obtained are rather tentative, both because of the approximation of the method and because of the natural variation in the size of the initial heterogeneities in  the samples. Therefore, these areas highlighted in yellow do not have a clear boundary in the plots. In Fig.\,\ref{F4} the crosses indicate the equilibrium parameters of the Al--Y melts whose time viscosity dependencies of relaxation are shown in Fig.\,\ref{fig1}. These parameters are in the determined above non-equilibrium area. Basing on our deductions, one can claim that the slow relaxation can be observed only in this area.

\section{Conclusions}

In conclusion, let us summarize our work. The main its statement is the origin for the slow relaxation in intermetallic melts is the Gibbs function nonlinearity happens as a result of the stoichiometric compound existence near considered liquid state in the phase diagram. This nonlinearity generates the thermodynamic instability similar to one taking place at spinodal decomposition. The difference of the thermodynamic instability in the considered relaxation processes from the one in the spinodal decomposition is  in the last one the thermodynamic instability leads to the development of heterogeneous structures, while as in the considered slow relaxation processes the thermodynamic instability leads to the slowing down of the relaxation of the initially heterogeneous structure to homogeneous state. Thus, the indispensable condition for manifestation of the slow relaxation processes is the initially non-homogenous.

The theoretical description of the relaxation processes in the heterogeneous melts with stoichiometric inclusions is carried out in terms of the Cahn--Hilliard equation with using of the combined Gibbs potential. We have shown the possibility of appears of long-life local instability in initially inhomogeneous binary melts. For the Al--Y and Al--Yb melts, the area in the phase diagram where this is possible is determined.
These results agree with conclusions presented earlier in \cite{Son}, where the width of the corresponding fluctuation region was estimated with the Ginsburg--Levanyuk criterion and also constituted up to several hundred degrees.
However, the considered local thermodynamic instability of initially heterogeneous melt, much slowing down its relaxation, does not still prevent the thermodynamic equilibrium with time. This time is defined by the scale of the initial inhomogeneous. This inhomogeneous scale corresponding to the experimentally observed relaxation time, $\tau\sim 10^4$ s, was estimated. It agrees with the experimentally measured values, $\xi\sim 10^{-5}$--$10^{-6}$ m.

The presented results have character of estimation. Of course, for more precise description of the slow relaxation in detail, one should solve the full system of the evolution equations for the phase field, concentration, and temperature. However, the results allow understanding the nature of slow and non-monotonic relaxation processes observed in some intermetallic melts after melting.

\section*{Acknowledgments}

The work was supported by the Russian Science Foundation under the project No. 21-13-00202.


\begin{thebibliography}{00}

\bibitem{Son} L. Son, M. Vasin, V. Sidorov, G. Rusakov, Journal of Alloys and Compounds, 785 (2019) 1279.
\bibitem{1} V.\,M.\,Zamiatin, B.\,A.\,Baum, A.\,A.\,Mezenina, et al., Rasplavi (in Russ.) 5 (2010) 19.
\bibitem{2} V.\,I.\,Lad'yanov , A.\,L.\,Bel'tyukov , S.\,G.\,Menshikova, et al., Physics and Chemistry of Liquids 46 (2008) 71.
\bibitem{3} V.\,I.\,Ladyanov, S.\,G.\,Menshikova, M.\,G.\,Vasin, et al.,  Bulletin of the Russian academy of sciences. Physics 75 (2011) 11
\bibitem{4}  S.\,G.\,Menshikova, A.\,L.\,Bel'tyukov, V.\,I.\,Lad'yanov, Journal of Advanced Materials 13 (2011) 533.
\bibitem{5}  A.\,L.\,Bel'tyukov, S.\,G.\,Menshikova, M.\,G.\,Vasin, et al., Rasplavi (in Russ.) 1 (2015) 3.
\bibitem{PB} P.\,S.\,Popel, B.\,A.\,Baum,  Izvestija AN SSSR. Metally 5 (1986) 47.
\bibitem{S1} I.\,V.\,Gavrilin, Izvestija AN SSSR. Metally (in Russ.) 2 (1985) 66.
\bibitem{S2} U.\,Dahlborg, M.\,Calvo-Dahlborg, P.\,S.\,Popel, V.\,E.\,Sidorov, Eur. Phys. J. B 14 (2000) 639.
\bibitem{Gasser} K.\,Khalouk, M.\,Mayoufi, J.\,G.\,Gasser,  Philosophical Magazine 90 (2010) 2695.
\bibitem{VMI} M.\,G.\,Vasin, S.\,G.\,Menshikova, M.\,D.\,Ivshin, Physica A 449 (2016) 64.
\bibitem{BLS} A.\,L.\,Beltyukov, B.\,A.\,Rusanov, D.\,A.\,Yagodin, A.\,I.\,Rusanova, E.\,V.\,Sterkhov, L.\,D.\,Son, V.\,I.\,Lad'yanov,
Solid State Communications, 360, (2023) 115044.
\bibitem{LV} V.\,G.\,Lebedev, A.\,A.\,Obukhov, and M.\,G.\,Vasin, Journal of Non-Crystalline Solids 505 (2019) 414.
\bibitem{RMET} S.\,G.\,Menshikova, I.\,G.\,Shirinkina, I.\,G.\,Brodova, and V.\,V.\,Brazhkin, Russian metallurgy (Metally) 2 (2019) 135.
\bibitem{CH} J.\,W.\,Cahn, J.\,E.\,Hilliard, J. Chem. Phys. 28 (1958) 258.
\bibitem{SS} V.\,P.\,Skripov, A.\,V.\,Skripov, Sov. Phys. Usp. 22 (1979) 389.
\bibitem{Elder} N.~Provatas, K.~Elder, \textit{Phase--Field Methods in Materials Science and Engineering}, Wiley-VCH Weinheim, (2010).
\bibitem{Flem} M.~C.~Flemings, \textit{Solidification processing}, McGraw-Hill,(1974) 364.
\bibitem{nims} URL: http://cpddb.nims.go.jp/cpddb/periodic.htm  25.09.2018.







\end{thebibliography}
\end{document}